\documentclass{article}
\usepackage{times,fancyhdr}
\usepackage{cite}
\usepackage{graphicx}

\title{The colour octet representation \\of the non-forward BFKL Green function\footnote{Preprint numbers: LPN11-27, IFT-UAM/CSIC-11-32, FTUAM-11-47}} 
\author{G. Chachamis$^1$, A. Sabio Vera$^2$ \\ 
\\
$^1$ Paul Scherrer Institut, CH-5232 Villigen PSI, Switzerland\\
\\
$^2$ Instituto de F{\' \i}sica Te{\' o}rica UAM/CSIC, Nicol{\'a}s Cabrera 15, and \\ 
Universidad Aut{\' o}noma de Madrid, E-28049 Madrid, Spain} 

\begin{document} 

\pagestyle{fancy}
\fancyhead{}
\fancyhead[EC]{G. Chachamis, J. D. Madrigal, A. Sabio Vera, P. Stephens}
\fancyhead[EL,OR]{\thepage}
\fancyhead[OC]{The colour octet representation of the non-forward BFKL Green function}
\fancyfoot{} 
\renewcommand\headrulewidth{0.5pt}
\addtolength{\headheight}{2pt} 

\maketitle 

We factorize the infrared divergences of the non-forward BFKL Green function for a general $t$-channel 
projection of the color quantum numbers and study the 
properties of the infrared finite remainder in the case of color octet exchange.  
The octet Green function is compared with the singlet case for different values of the momentum transfer. 
The octet Green function plays an important role in the calculation of the finite 
remainder of the two-loop six-point MHV planar amplitude in ${\cal N} = 4$ SYM as it was demonstrated 
by Bartels, Lipatov and one of us in~\cite{Bartels:2008ce,Bartels:2008sc}. 
A comparison with regularizations preserving conformal invariance at large momentum transfer is shown.

\section{Introduction}

The Balitsky-Fadin-Kuraev-Lipatov (BFKL) formalism\cite{BFKL1,BFKL2,BFKL3} has been used to investigate the properties of scattering amplitudes in Quantum Chromodynamics (QCD) and ${\cal N} = 4$ supersymmetric 
Yang-Mills theory (MSYM) in certain kinematic regions (multi-Regge (MRK) and quasi-multi-Regge kinematics) where logarithms of the center-of-mass energy are enhanced. In 
the leading and next-to-leading logarithmic~\cite{Fadin:1998py,Ciafaloni:1998gs} 
approximations the elastic $2 \to 2$ and inelastic 
$2 \to 2 + n$ amplitudes have an iterative structure given in terms of reggeized gluon propagators 
and ``squared" effective emission vertices. This iterative structure can be written for a general color representation in terms of a gluon Green function obeying a linear Bethe-Salpeter equation. The so-called ``hard" or BFKL pomeron corresponds to the projection in the $t$-channel of a singlet in color space. This is the most widely used representation since it is 
related to physical observables such as hadron structure functions at small values of Bjorken $x$ 
in deep inelastic scattering or inclusive dijet production with a significant rapidity separation at the Large Hadron Collider~\cite{Vera:2006un,Bartels:2006hg,Vera:2007kn,Vera:2007dr,Chachamis:2009ks,Chachamis:2011rw,arXiv:1106.6172,arXiv:1110.5830,arXiv:1110.6741}.

The BFKL resummation program has also been very useful from a more formal point of view. It was in a generalized leading logarithmic approximation, iterating the BFKL kernel in the $s$-channel, where the Bartels-Kwiecinski-Praszalowicz (BKP) equation was 
proposed~\cite{Bartels:1980pe,Kwiecinski:1980wb} and found to have a hidden 
integrability~\cite{Lipatov:1985uk,Lipatov:1990zb,Lev1}, being equivalent to a periodic spin chain of a XXX Heisenberg 
ferromagnet~\cite{Lev2,Lipatov:1994xy,Faddeev:1994zg}. This was the first example of the existence of integrable systems in QCD. 
A similar integrable spin chain, an open one this time, has recently been found by Lipatov in kinematical regions of $n$-point maximally helicity violating (MHV) and 
planar ($N_c \to \infty$) amplitudes in MSYM where Mandelstam cut contributions are maximally enhanced~\cite{Lipatov:2009nt,Bartels:2011nz}. Corrections to the Bern-Dixon-Smirnov (BDS) iterative ansatz~\cite{Bern:2005iz} for this class of amplitudes were found 
in MRK by Bartels, Lipatov and one of the authors in the six-point amplitude at 
two loops in~\cite{Bartels:2008ce,Bartels:2008sc}. These corrections have been understood as part of the finite remainder~\cite{Drummond:2007au,Drummond:2008aq,Bern:2008ap} to the amplitude which corresponds to the anomalous contribution of a conformal Ward identity. A very compact formula for the amplitude can be found in 
terms of Goncharov polylogarithms~\cite{DelDuca:2010zg,Goncharov:2010jf} which is in 
agreement with the corresponding Regge limits~\cite{Lipatov:2010ad,Bartels:2010tx,Bartels:2011xy} (also at three loop 
level~\cite{Dixon:2011pw}). 
Very recently, Romagnoni and one of us have found a novel relation between the BFKL equation in the forward limit and  
the sl(2) invariant XXX spin -1/2 chain~\cite{arXiv:1111.4553}.

The octet gluon Green function is a fundamental ingredient of the finite remainder of scattering amplitudes with arbitrary 
number of external legs and internal loops. We investigate its all-orders structure in detail in the present work. In Section 2 
we present an iterative representation for the Green function with a general color projection in the $t$-channel written 
in rapidity and transverse momentum space. We show how its 
infrared divergencies can be factorized in a simple form, leaving an iterative finite remainder that we investigate  in detail in Section 3. Here we use advanced Monte Carlo techniques~\cite{code}  to study the exclusive information 
present in this Green function in both the singlet and octet projections. There appears faster convergence in the octet solution with a similar behavior in both forward and 
non-forward solutions. The collinear regions, related to the calculation of anomalous dimensions, are also investigated. In Section 4 we discuss the factorization of the infrared divergencies used by Bartels, Lipatov and one of us in Ref.~\cite{Bartels:2008ce,Bartels:2008sc} where they are extracted introducing a momentum transfer dependent gluon Regge trajectory. A similar regularization is used by Fadin and Lipatov in Ref.~\cite{arXiv:1111.0782}, which is also discussed here. We show the relation among these 
representations and discuss the change in the role played by the $n=0$ and $n=1$ conformal spins.
In Section 5 we present our Conclusions and scope for future work. 

\section{The Green function for a general colour projection}

The non-forward BFKL equation for a general projection of the colour quantum numbers in the $t$-channel can be found, at leading and next-to-leading order, 
in Ref.~\cite{arXiv:1111.0782,Fadin:2005zj}. In the present work we are interested in comparing 
the gluon Green function in the singlet case with that of the octet projection. The only 
difference between both solutions is in the ``real emission" part of the kernel, which 
in the color octet case carries an extra factor of $1/2$ with respect to the singlet case. 
In the octet case this spoils the complete cancellation of infrared divergencies present in the singlet, or pomeron, projection.  

To show that the extra infrared divergencies that appear in the non-singlet 
representations can be written as a simple overall factor in the gluon Green 
function, we regularize half of the divergencies in the gluon Regge trajectory 
using dimensional $D=4-2 \,\epsilon$ regularization, while the remaining ones are 
treated using a mass parameter $\lambda$, which is also used to regularize the 
phase space integral of the ``real emission" sector. The dependence on $\lambda$ 
will cancel out while the dependence on $\epsilon$ will remain in the factorized term. We use the notation ${\bf q}_i' \equiv {\bf q}_i - {\bf q}$, where ${\bf q}$ is the 
momentum transfer and all two--dimensional vectors are represented in 
bold. 

In this way, the non-forward BFKL equation for a kernel in a 
general color group representation ${\cal R}$ in the $t$--channel can 
be written as
\begin{eqnarray}
\Bigg\{\omega + (c_{\cal R}-1) \frac{{\bar \alpha}_s}{2} 
\left[\frac{2}{\epsilon}
-\log{\left({{\bf q}_1^2 \over \mu^2}\right)}
-\log{\left({{\bf q}_1'^2  \over \mu^2}\right)}
\right] &&\nonumber\\
&&\hspace{-7cm}+c_{\cal R} {{\bar \alpha}_s \over 2}
\left[\log{\left({{\bf q}_1^2 \over \lambda^2}\right)}
+\log{\left({{\bf q}_1'^2  \over \lambda^2}\right)}
\right]\Bigg\} {\cal G}_\omega \left({\bf q}_1,{\bf q}_2;{\bf q}\right) ~=~ 
\delta^{(2)} \left({\bf q}_1-{\bf q}_2\right)\nonumber\\
&&\hspace{-7cm}+c_{\cal R} \int {d^2 {\bf k} \over \pi {\bf k}^2} 
\theta \left({\bf k}^2 - \lambda^2\right)
{{\bar \alpha}_s \over 2} 
\Bigg[1+{{\bf q}_1'^2 ({\bf q}_1+{\bf k})^2 - {\bf q}^2 {\bf k}^2 \over 
({\bf q}_1'+{\bf k})^2 {\bf q}_1^2}\Bigg]
{\cal G}_\omega \left({\bf q}_1+{\bf k},{\bf q}_2;{\bf q}\right).
\label{BFKL1}
\end{eqnarray}
Note that compared to~\cite{Fadin:2005zj} all we have done is to add and 
subtract two gluon trajectories with a $c_{\cal R}$ factor in front of them. 
The solution of this equation is independent of $\lambda$ in the 
$\lambda \rightarrow 0$ limit. $\mu$ is the renormalization scale in the 
$\overline{\rm MS}$ scheme. As we pointed out before, in the singlet representation 
$c_1 = 1$ and all infrared divergencies cancel out. In the 
octet representations $c_{8_a} = c_{8_s}= 1/2$ and the divergent behaviour of the 
gluon trajectory times $(c_{\cal R}-1)$ remains. Other representations, which we will not investigate here, 
are $c_{10} = c_{\overline{10}} = 0$ and $c_{27} = - c_{N_c>3} = -1/(4 N_c)$.

We solve Eq.~(\ref{BFKL1}) first introducing the following notation:
\begin{eqnarray}
\left\{\omega - \omega^{(\epsilon;\lambda)} ({\bf q}_1;{\bf q})\right\} 
{\cal G}_\omega \left({\bf q}_1,{\bf q}_2;{\bf q}\right) &=& 
\delta^{(2)} \left({\bf q}_1-{\bf q}_2\right)\nonumber\\
&&\hspace{-4cm}+c_{\cal R} \int {d^2 {\bf k} \over \pi {\bf k}^2} 
\theta \left({\bf k}^2 - \lambda^2\right)
\xi \left({\bf q}_1,{\bf k};{\bf q}\right)
{\cal G}_\omega \left({\bf q}_1+{\bf k},{\bf q}_2;{\bf q}\right).
\end{eqnarray}
This can now be iterated and, performing the Mellin transform, 
\begin{eqnarray}
{\cal F} \left({\bf q}_1,{\bf q}_2;{\bf q};{\rm Y}\right) &=& 
\int {d \omega \over 2 \pi i} e^{\omega {\rm Y}} 
{\cal G}_\omega \left({\bf q}_1,{\bf q}_2;{\bf q}\right),
\end{eqnarray}
we finally obtain
\begin{eqnarray}
{\cal F} \left({\bf q}_1,{\bf q}_2;{\bf q};{\rm Y}\right) &=& 
\exp{\left\{\omega^{(\epsilon;\lambda)} ({\bf q}_1;{\bf q}){\rm Y}\right\}} 
\Bigg\{ 
\delta^{(2)} \left({\bf q}_1-{\bf q}_2\right) \nonumber\\
&&\hspace{-3.6cm}+\sum_{n=1}^\infty \prod_{i=1}^n c_{\cal R} \int 
{d^2 {\bf k}_i \over \pi {\bf k}_i^2} \theta ({\bf k}_i^2-\lambda^2) 
\, \xi \left({\bf q}_1+\sum_{l=1}^{i-1}{\bf k}_l,{\bf k}_i;{\bf q}
\right) \delta^{(2)} \left({\bf q}_1+\sum_{l=1}^n {\bf k}_l-{\bf q}_2\right)
\nonumber\\
&&\hspace{-3.6cm}\times
\int_0^{y_{i-1}} d y_i \exp{\left\{\left[\omega^{(\epsilon;\lambda)} 
\left({\bf q}_1+\sum_{l=1}^i {\bf k}_l;{\bf q}\right)
-\omega^{(\epsilon;\lambda)} 
\left({\bf q}_1+\sum_{l=1}^{i-1} {\bf k}_l;{\bf q}\right)\right]y_i\right\}}
\Bigg\},
\label{itesol}
\end{eqnarray}
where the rapidity $y_0 \equiv {\rm Y} \sim \log{s}$ is given by the logarithm of 
the center of mass energy $\sqrt{s}$. 

The $\lambda$ and $1/\epsilon$ dependence in the Reggeon propagators ({\it i.e.}, all the exponentials  
in Eq.~(\ref{itesol})) appear only in the initial multiplicative factor since
\begin{eqnarray}
\omega^{(\epsilon;\lambda)} 
\left({\bf q}_1+\sum_{l=1}^i {\bf k}_l;{\bf q}\right)
-\omega^{(\epsilon;\lambda)} 
\left({\bf q}_1+\sum_{l=1}^{i-1} {\bf k}_l;{\bf q}\right) &=& \nonumber\\
&&\hspace{-8cm}-{{\bar \alpha}_s \over 2}
\log{\left(\left({\bf q}_1+\sum_{l=1}^i{\bf k}_l\right)^2
\left({\bf q}_1'+\sum_{l=1}^i{\bf k}_l\right)^2 \over 
\left({\bf q}_1+\sum_{l=1}^{i-1}{\bf k}_l\right)^2
\left({\bf q}_1'+\sum_{l=1}^{i-1}{\bf k}_l\right)^2 \right)}
\end{eqnarray}
for any $t$--channel colour representation.

This implies that the gluon Green function can then be written in the form
\begin{eqnarray}
{\cal F} \left({\bf q}_1,{\bf q}_2;{\bf q};{\rm Y}\right) &=& 
e^{(1-c_{\cal R}) \frac{{\bar \alpha}_s}{\epsilon}{\rm Y}}\left({\sqrt{{\bf q}_1^2 {\bf q}_1'^2} \over \mu^{2(1-c_{\cal R})}\lambda^{2 c_{\cal R}}}\right)^{-{{\bar \alpha}_s}{\rm Y}}\Bigg\{\delta^{(2)} \left({\bf q}_1-{\bf q}_2\right) \nonumber\\&&\hspace{-3.2cm}+\sum_{n=1}^\infty \prod_{i=1}^n c_{\cal R} \int 
{d^2 {\bf k}_i \over \pi {\bf k}_i^2} \theta ({\bf k}_i^2-\lambda^2) 
\,  {{\bar \alpha}_s \over 2} 
\Bigg(1+{\left({\bf q}_1'+\sum_{l=1}^{i-1}{\bf k}_l\right)^2 ({\bf q}_1+\sum_{l=1}^{i}{\bf k}_l)^2 - {\bf q}^2 {\bf k}_i^2 \over 
({\bf q}_1'+\sum_{l=1}^{i}{\bf k}_l)^2 \left({\bf q}_1+\sum_{l=1}^{i-1}{\bf k}_l\right)^2}\Bigg)\nonumber\\
&&\hspace{-3cm}\times
\int_0^{y_{i-1}} \hspace{-0.4cm} d y_i \left(\left({\bf q}_1+\sum_{l=1}^{i-1}{\bf k}_l\right)^2
\left({\bf q}_1'+\sum_{l=1}^{i-1}{\bf k}_l\right)^2 \over 
\left({\bf q}_1+\sum_{l=1}^{i}{\bf k}_l\right)^2
\left({\bf q}_1'+\sum_{l=1}^{i}{\bf k}_l\right)^2 \right)^{{{\bar \alpha}_s \over 2}y_i}\hspace{-0.6cm} \delta^{(2)} \left({\bf q}_1+\sum_{l=1}^n {\bf k}_l-{\bf q}_2\right) \Bigg\}.
\end{eqnarray}
The forward limit, ${\bf q}=0$, is particularly simple:
\begin{eqnarray}
{\cal F} \left({\bf q}_1,{\bf q}_2;{\rm Y}\right) &=& 
e^{(1-c_{\cal R}) \frac{{\bar \alpha}_s}{\epsilon}{\rm Y}}
\left({{\bf q}_1^2 \over \mu^{2(1-c_{\cal R})}\lambda^{2 c_{\cal R}}}\right)^{-{{\bar \alpha}_s}{\rm Y}}
\nonumber\\
&&\hspace{-3cm}\times\Bigg\{\delta^{(2)} \left({\bf q}_1-{\bf q}_2\right) +\sum_{n=1}^\infty \prod_{i=1}^n {{\bar \alpha}_s} c_{\cal R} \int 
{d^2 {\bf k}_i \over \pi {\bf k}_i^2} \theta ({\bf k}_i^2-\lambda^2)\nonumber\\
&&\hspace{-1cm}\times
\int_0^{y_{i-1}} \hspace{-0.4cm} d y_i \left(\left({\bf q}_1+\sum_{l=1}^{i-1}{\bf k}_l\right)^2 \over 
\left({\bf q}_1+\sum_{l=1}^{i}{\bf k}_l\right)^2 \right)^{{{\bar \alpha}_s}y_i}\hspace{-0.6cm} \delta^{(2)} \left({\bf q}_1+\sum_{l=1}^n {\bf k}_l-{\bf q}_2\right) \Bigg\}.\end{eqnarray}
In the case of colour singlet $c_1 = 1$ the solution is well-known:
\begin{eqnarray}
{\cal F} \left({\bf q}_1,{\bf q}_2;{\rm Y}\right) &=& 
\frac{1}{\pi \sqrt{{\bf q}_1^2 {\bf q}_2^2}}
\sum_{n=-\infty}^\infty \int \frac{d\omega}{2 \pi i} 
\int \frac{d\gamma}{2 \pi i} 
\left(\frac{{\bf q}_1^2}{{\bf q}_2^2}\right)^{\gamma-\frac{1}{2}} 
\frac{e^{\omega {\rm Y}+i \, n \,\theta}}{\omega 
- {\bar \alpha}_s \chi_n (\gamma)},
\label{Analytic}
\end{eqnarray}
where
\begin{eqnarray}
\chi_n (\gamma) &=& 2 \Psi(1) 
- \Psi\left(\gamma + \frac{|n|}{2}\right)
- \Psi\left(1-\gamma + \frac{|n|}{2}\right)
\end{eqnarray}
is the eigenvalue of the BFKL kernel and $\Psi$ is the logarithmic 
derivative of 
Euler's Gamma function. $\theta$ is the azimuthal angle between the 
two--dimensional vectors ${\bf q}_1$ and ${\bf q}_2$. 

From now on we will focus on the description of the infrared finite remainder in the form
\begin{eqnarray}
{\cal H} \left({\bf q}_1,{\bf q}_2;{\bf q};{\rm Y}\right) &\equiv& 
{\cal F}\left({\bf q}_1,{\bf q}_2;{\bf q};{\rm Y}\right) 
\left(\frac{e^{\frac{1}{\epsilon}} \mu^2}{\sqrt{{\bf q}_1^2 {\bf q}_1'^2} } 
\right)^{{\bar \alpha}_s (c_{\cal R}-1){\rm Y} } \nonumber\\
&=& 
\left(
{\lambda^{2 } \over \sqrt{{\bf q}_1^2 {\bf q}_1'^2} }
\right)^{{c_{\cal R} {\bar \alpha}_s}{\rm Y}} 
\Bigg\{\delta^{(2)} \left({\bf q}_1-{\bf q}_2\right) \nonumber\\&&\hspace{-3.2cm}+\sum_{n=1}^\infty \prod_{i=1}^n c_{\cal R} \int 
{d^2 {\bf k}_i \over \pi {\bf k}_i^2} \theta ({\bf k}_i^2-\lambda^2) 
\,  {{\bar \alpha}_s \over 2} 
\Bigg(1+{\left({\bf q}_1'+\sum_{l=1}^{i-1}{\bf k}_l\right)^2 ({\bf q}_1+\sum_{l=1}^{i}{\bf k}_l)^2 - {\bf q}^2 {\bf k}_i^2 \over 
({\bf q}_1'+\sum_{l=1}^{i}{\bf k}_l)^2 \left({\bf q}_1+\sum_{l=1}^{i-1}{\bf k}_l\right)^2}\Bigg)\nonumber\\
&&\hspace{-3cm}\times
\int_0^{y_{i-1}} \hspace{-0.4cm} d y_i \left(\left({\bf q}_1+\sum_{l=1}^{i-1}{\bf k}_l\right)^2
\left({\bf q}_1'+\sum_{l=1}^{i-1}{\bf k}_l\right)^2 \over 
\left({\bf q}_1+\sum_{l=1}^{i}{\bf k}_l\right)^2
\left({\bf q}_1'+\sum_{l=1}^{i}{\bf k}_l\right)^2 \right)^{{{\bar \alpha}_s \over 2}y_i}\hspace{-0.6cm} \delta^{(2)} \left({\bf q}_1+\sum_{l=1}^n {\bf k}_l-{\bf q}_2\right) \Bigg\},
\label{H}
\end{eqnarray}
where we have removed the factor with the $\epsilon$ and $\mu$ dependence. We have checked that this function ${\cal H}$ is $\lambda$ independent for small 
$\lambda$. In the following section we will compare this infrared finite remainder 
${\cal H}$ for $c_{\cal R}=1 $ and $1/2$.

\section{Comparing the octet with the singlet gluon Green function}

\begin{figure}[htbp]
\hspace{-1.2cm}  \includegraphics[width=8cm,angle=0]{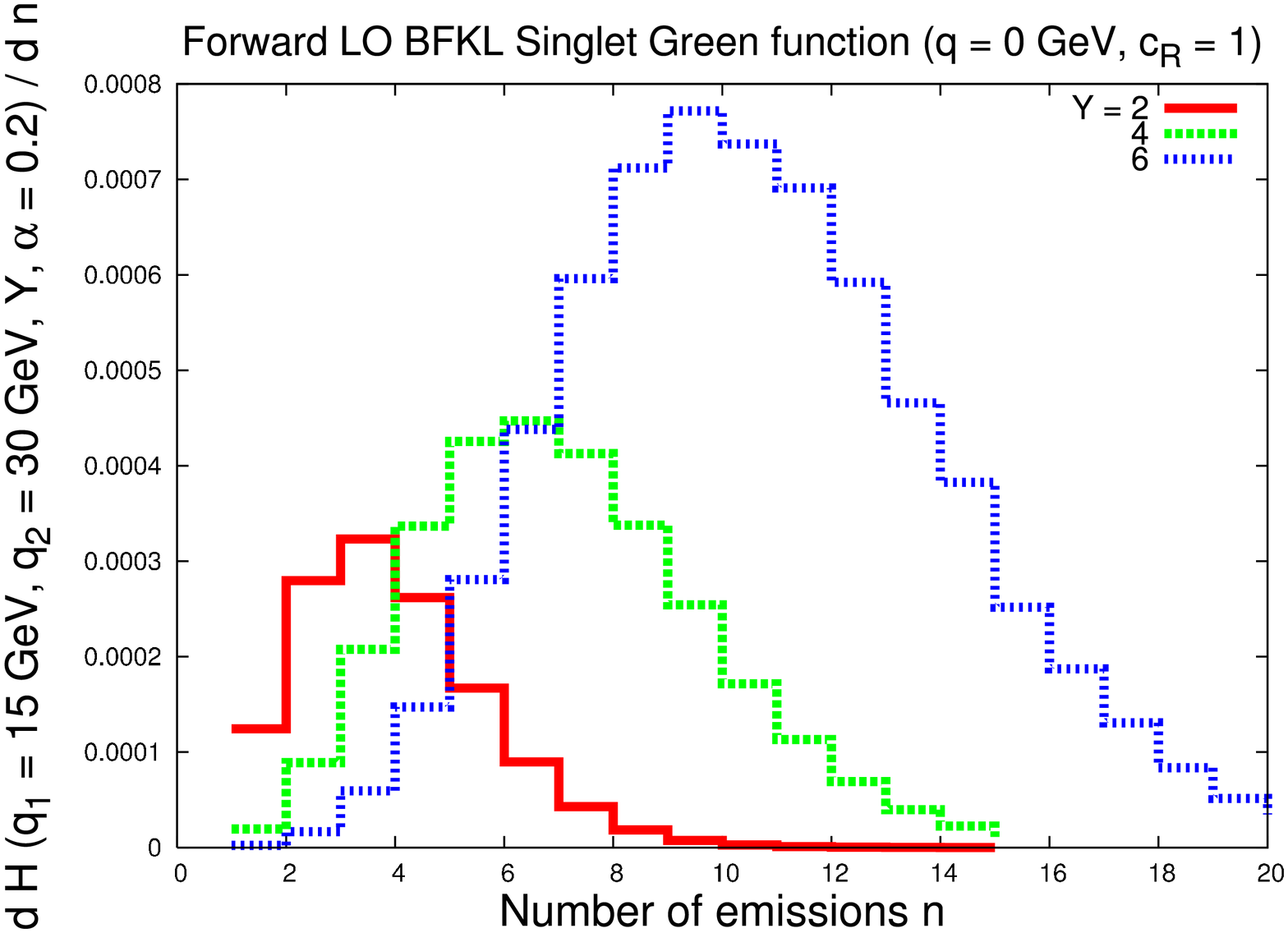}\includegraphics[width=8cm,angle=0]{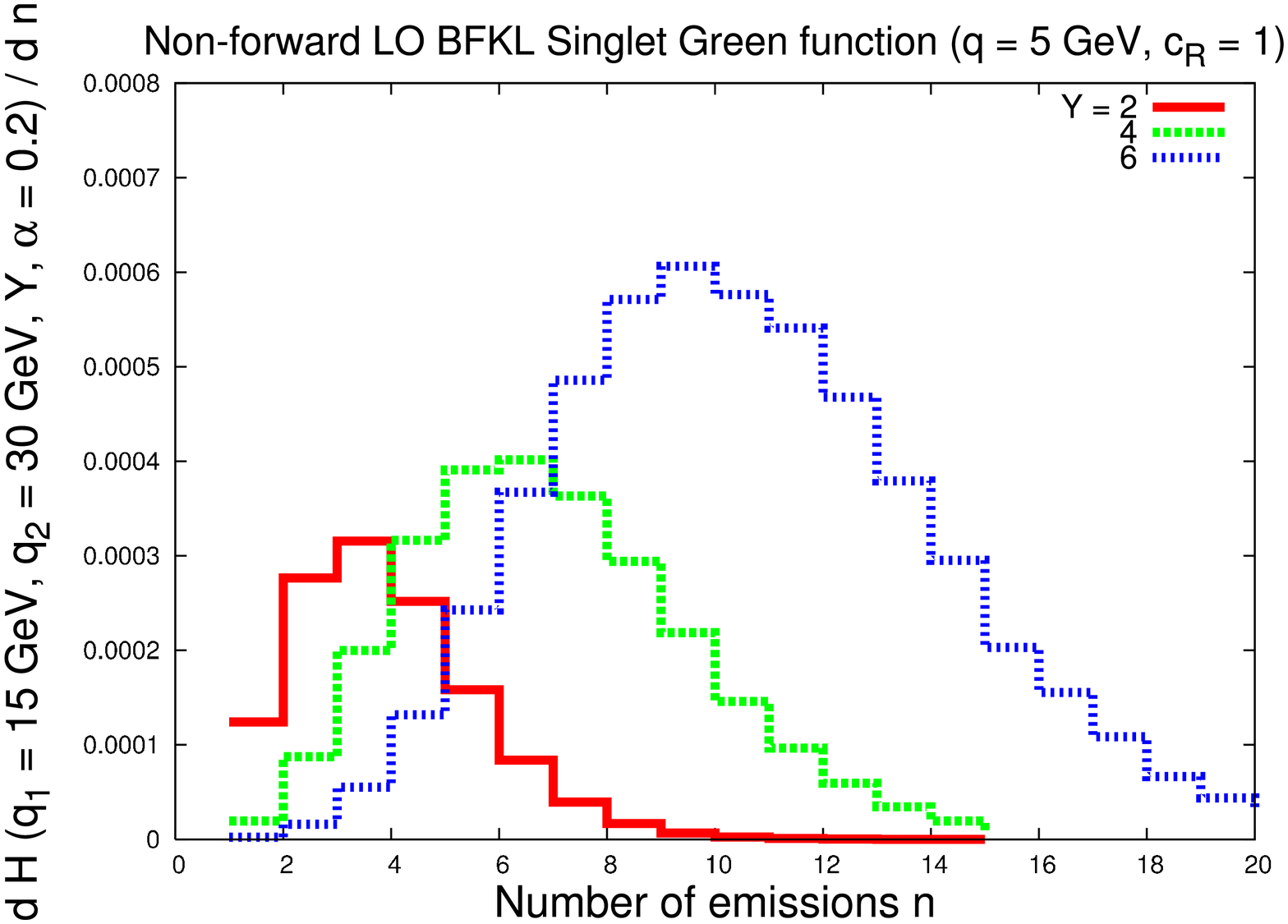}

\hspace{-1.2cm}  \includegraphics[width=8cm,angle=0]{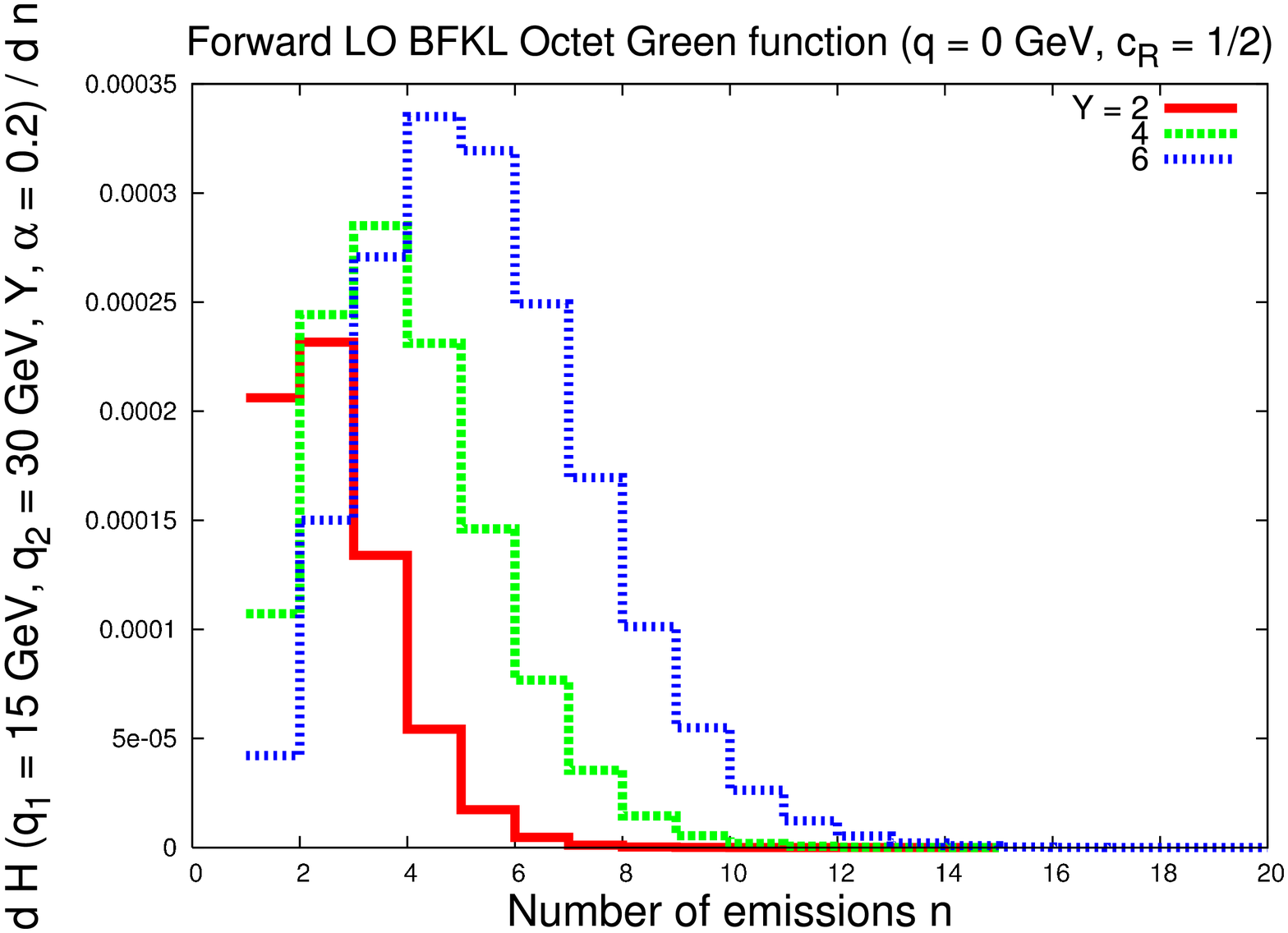}\includegraphics[width=8cm,angle=0]{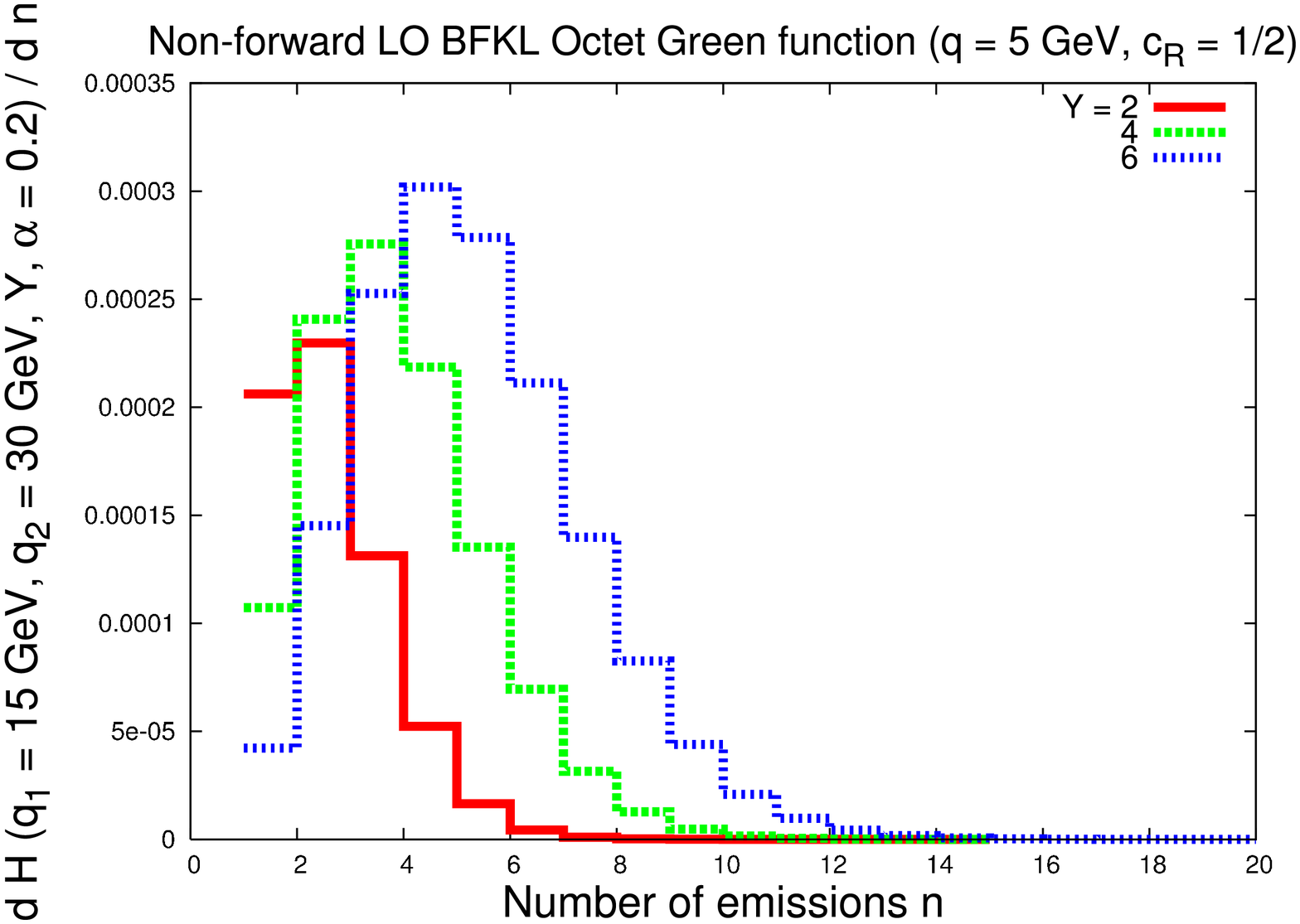}
    \caption{Distribution in the contributions to the BFKL gluon Green function 
    with a fixed number of iterations of the kernel, plotted for different values of the 
    center-of-mass energy, and a fixed ${\bar \alpha}_s = 0.2$.}
  \label{NumberOfTerms}
\end{figure}

An interesting question is to study the convergence of the sum defining the 
function ${\cal H}$ in Eq.~(\ref{H}). For a fixed value of Y and the coupling 
${\bar \alpha}_s$ a 
finite number of terms in the sum is needed to reach a good accuracy for the 
gluon Green function. As the value of the effective parameter 
${\bar \alpha}_s {\rm Y}$ 
gets larger the Green function is more sensitive to high multiplicity terms, following 
a Poissonian distribution as it can be seen in Fig.~\ref{NumberOfTerms}.  
In both plots at the top of the figure, which correspond to the singlet 
forward $q=0$ (left) and non-forward, with $q=5$ GeV (right), cases, we can see how the distribution in the number of iterations of the kernel gets broader for 
larger center-of-mass energies. Despite this there is good convergence, 
{\it e.g.}, for ${\rm Y} = 4$, $q_1 = 15$ GeV and $q_2 = 30$ GeV, it is enough to keep up to 16 terms in the sum to get a very accurate result. It is remarkable that 
the convergence is much better in the octet case where it is possible to get the Green function with a small number of terms, see Fig.~\ref{NumberOfTerms} (bottom). This can be qualitatively understood if we 
think of the ``real emission" terms as pushing the Green function to grow with $Y$. In the case of the singlet 
they carry the same coefficient as the contributions from the gluon Regge trajectories, which make the Green 
function to decrease with $Y$. For the octet this 
balance is broken in favor of the latter making the Green function to grow much slower with an effective 
reduction in the number of emissions. 

\begin{figure}[htbp]
\hspace{-1.2cm}  
\includegraphics[width=8cm,angle=0]{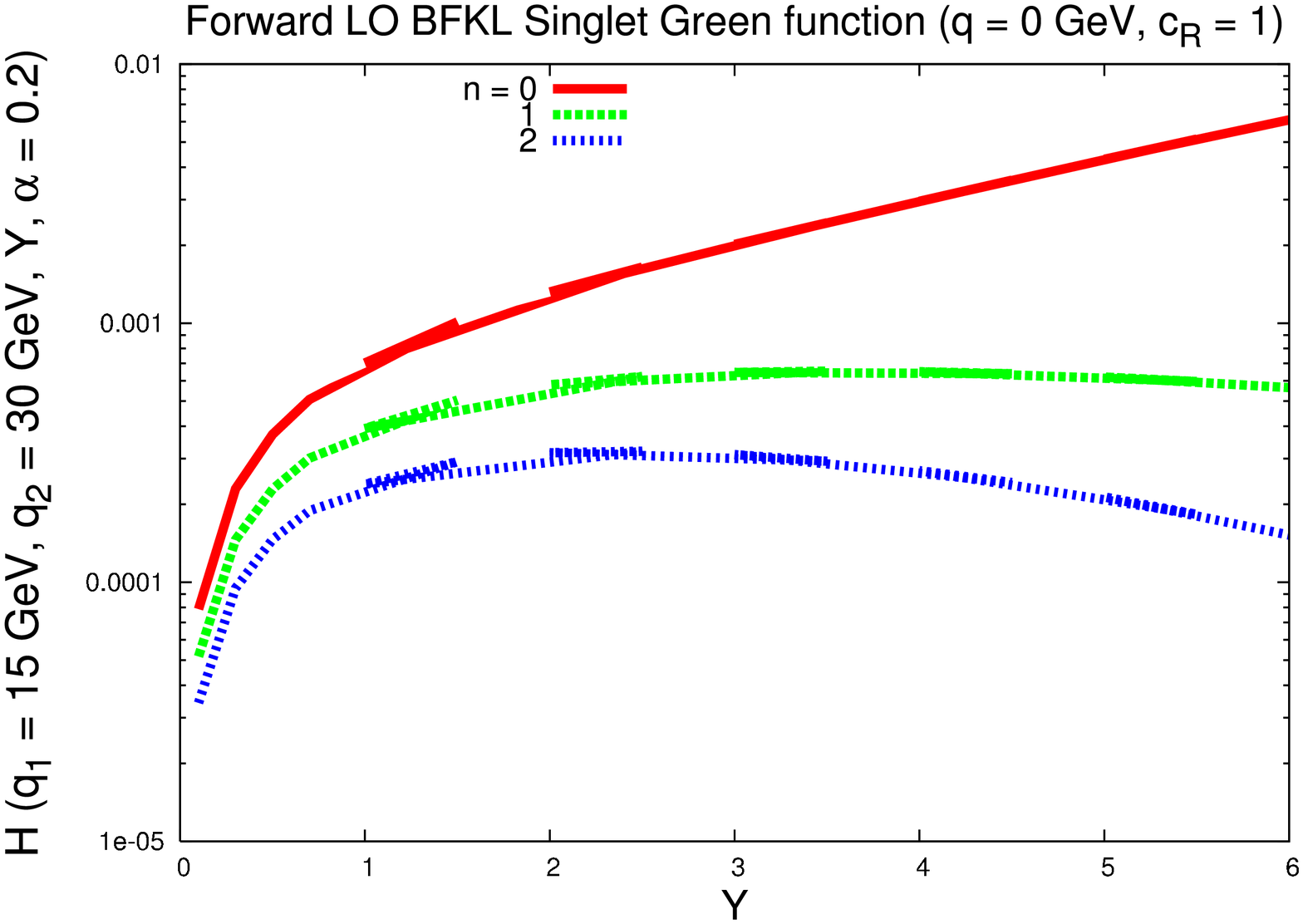}\includegraphics[width=8cm,angle=0]{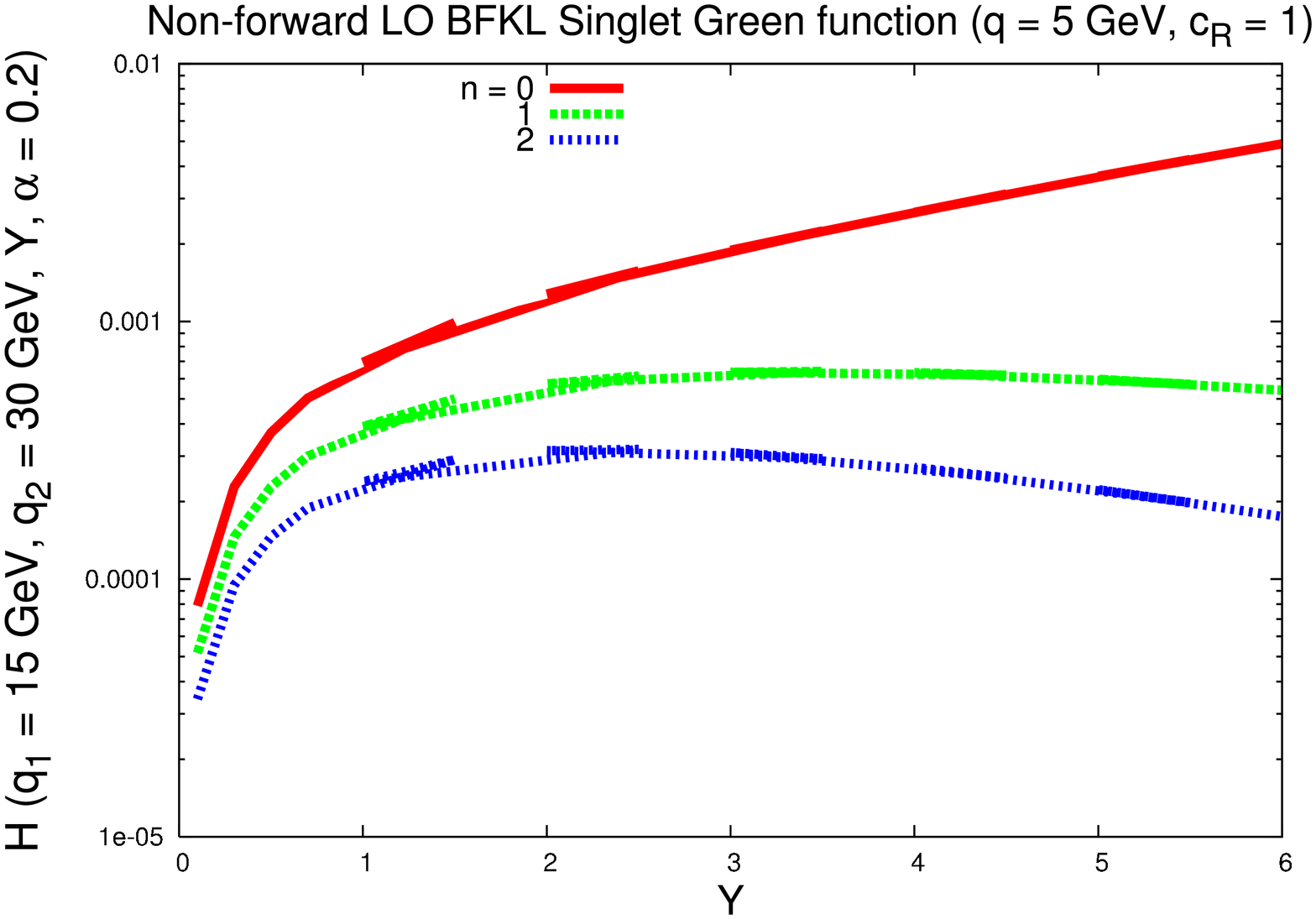}

\hspace{-1.2cm} \includegraphics[width=8cm,angle=0]{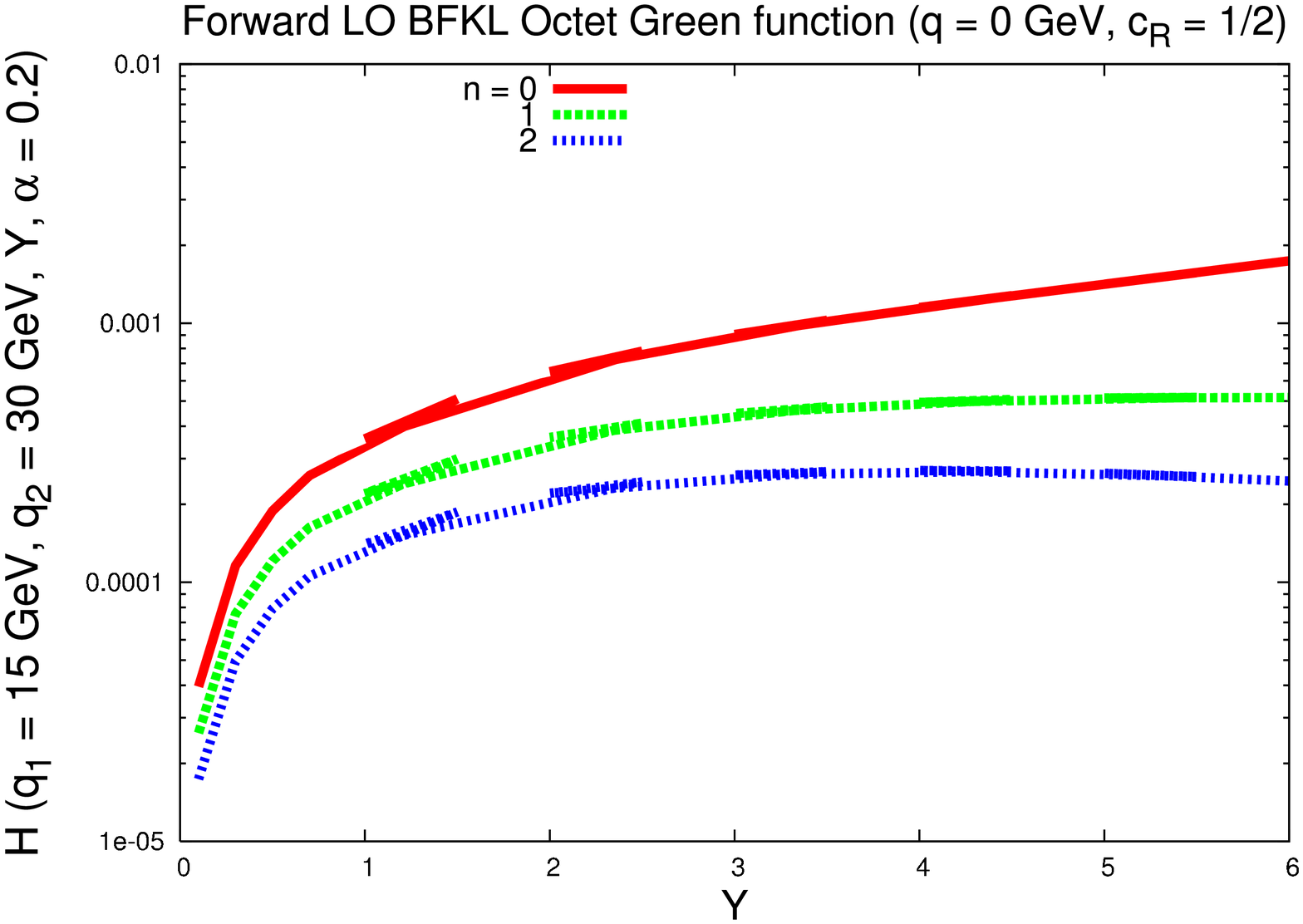}\includegraphics[width=8cm,angle=0]{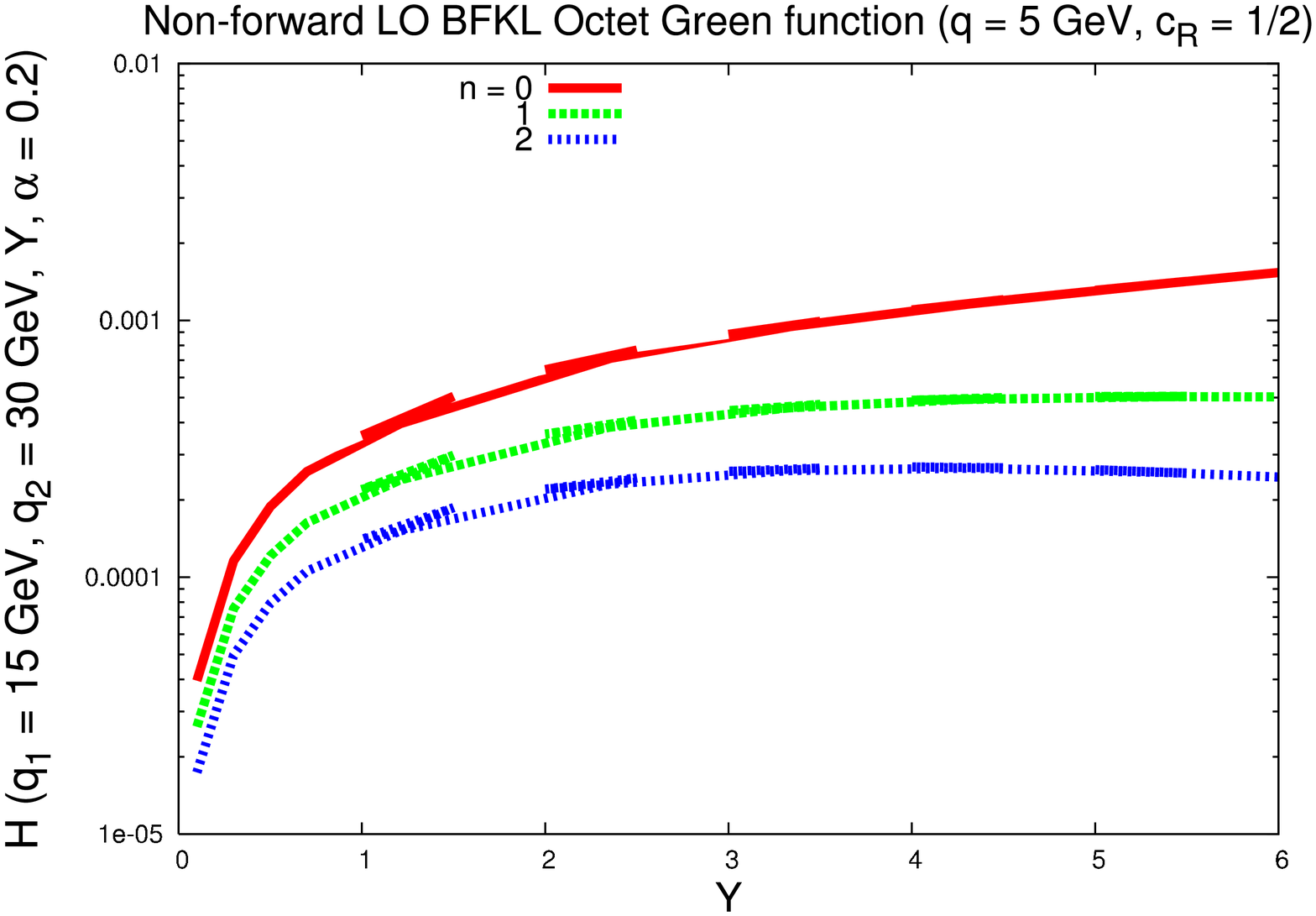}
\caption{Projection of the gluon Green function on different Fourier components in the 
azimuthal angle between the ${\bf q}_1$ and ${\bf q}_2$ transverse momenta.}
\label{Angles-n}
\end{figure}
It is also instructive to compare both solutions in terms of the different Fourier 
components in the azimuthal angle between the two momenta ${\bf q}_1$ and 
${\bf q}_2$. In the particular case of the singlet and forward case this expansion 
is given in Eq.~(\ref{Analytic}). A complete analysis for the singlet and octet 
projections is shown in Fig.~\ref{Angles-n}. There are no surprises since both 
solutions behave quite similarly, with the only rising component being the 
$n=0$ one. 
\begin{figure}[htbp]
\hspace{-1.2cm}  \includegraphics[width=8cm,angle=0]{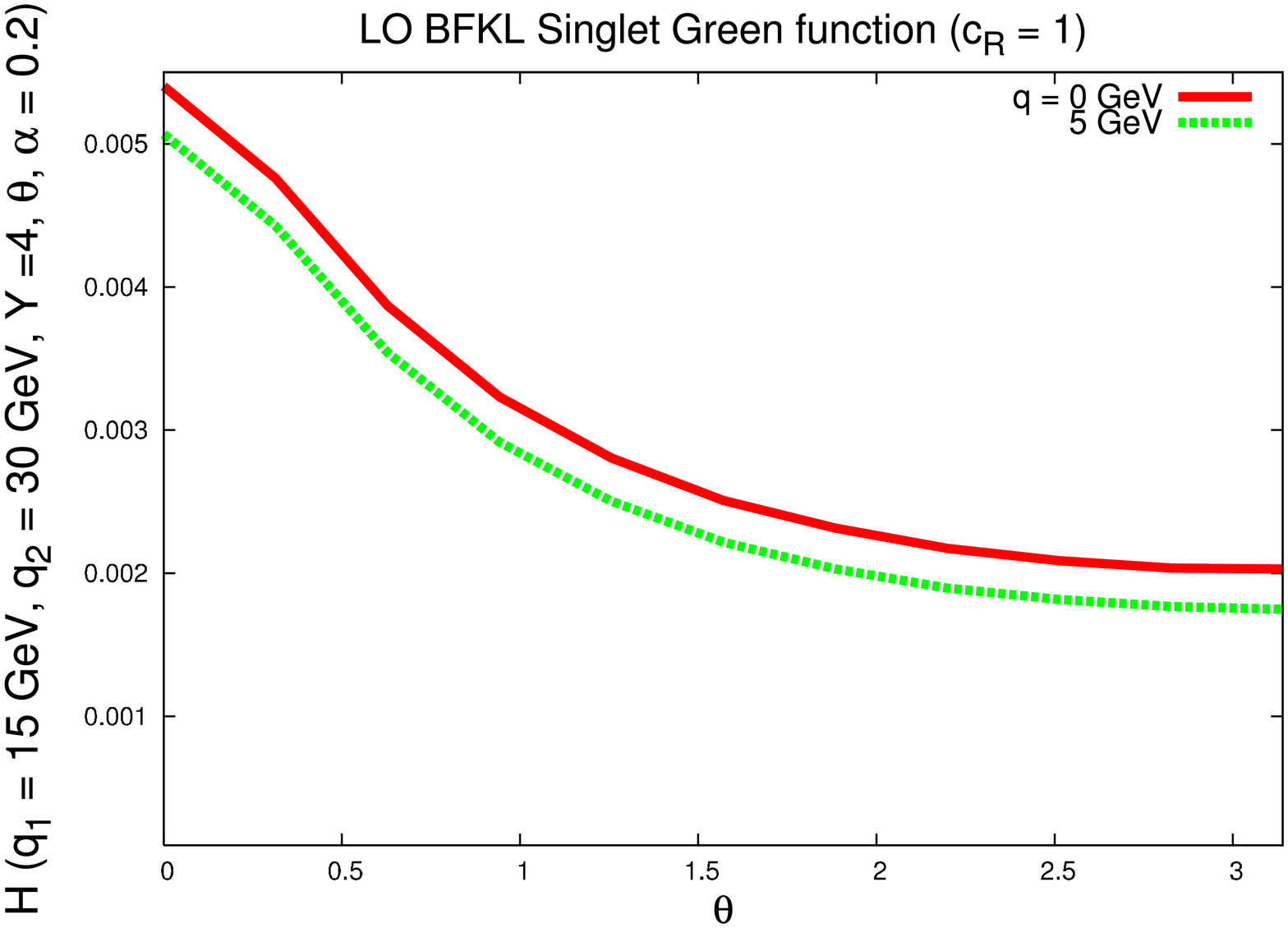}\includegraphics[width=8cm,angle=0]{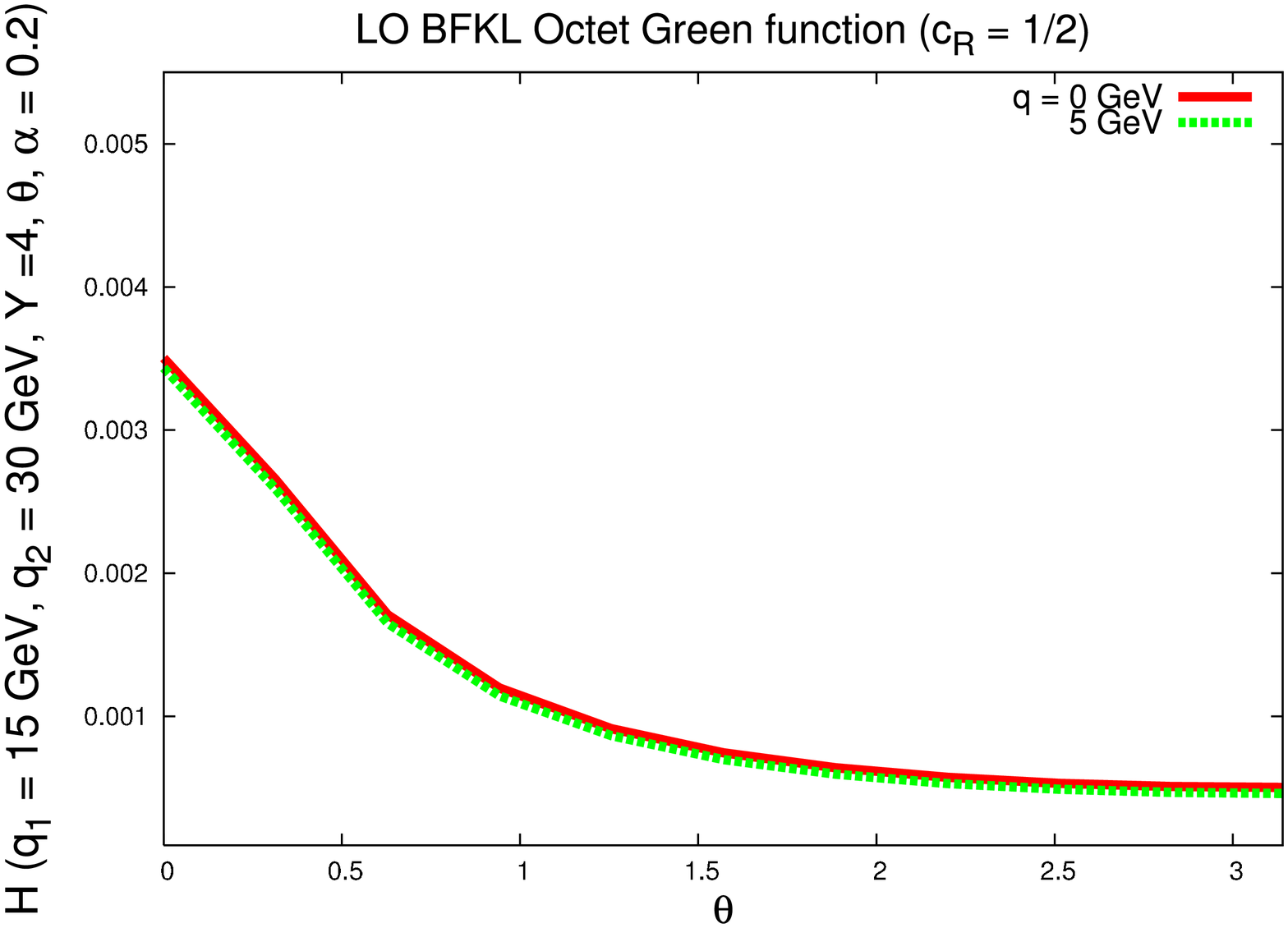}
\caption{Gluon Green function dependence for the full range in azimuthal angles.}  
\label{BFKLfullangle}
\end{figure}
It is possible to appreciate the similarity between both color projections in the full range of azimuthal angles in Fig.~\ref{BFKLfullangle}.

\begin{figure}[htbp]
\hspace{-1.2cm}  \includegraphics[width=8cm,angle=0]{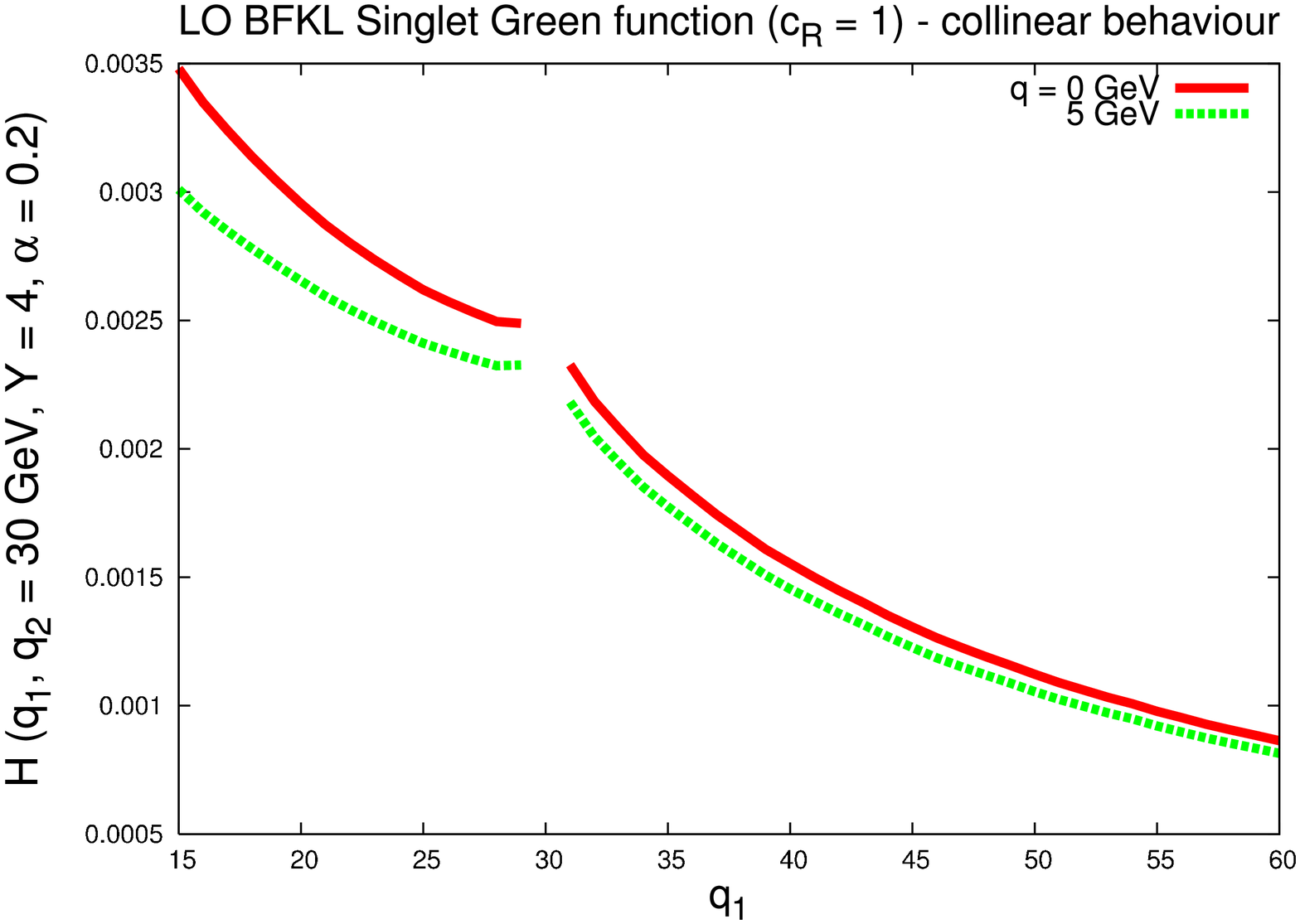}\includegraphics[width=8cm,angle=0]{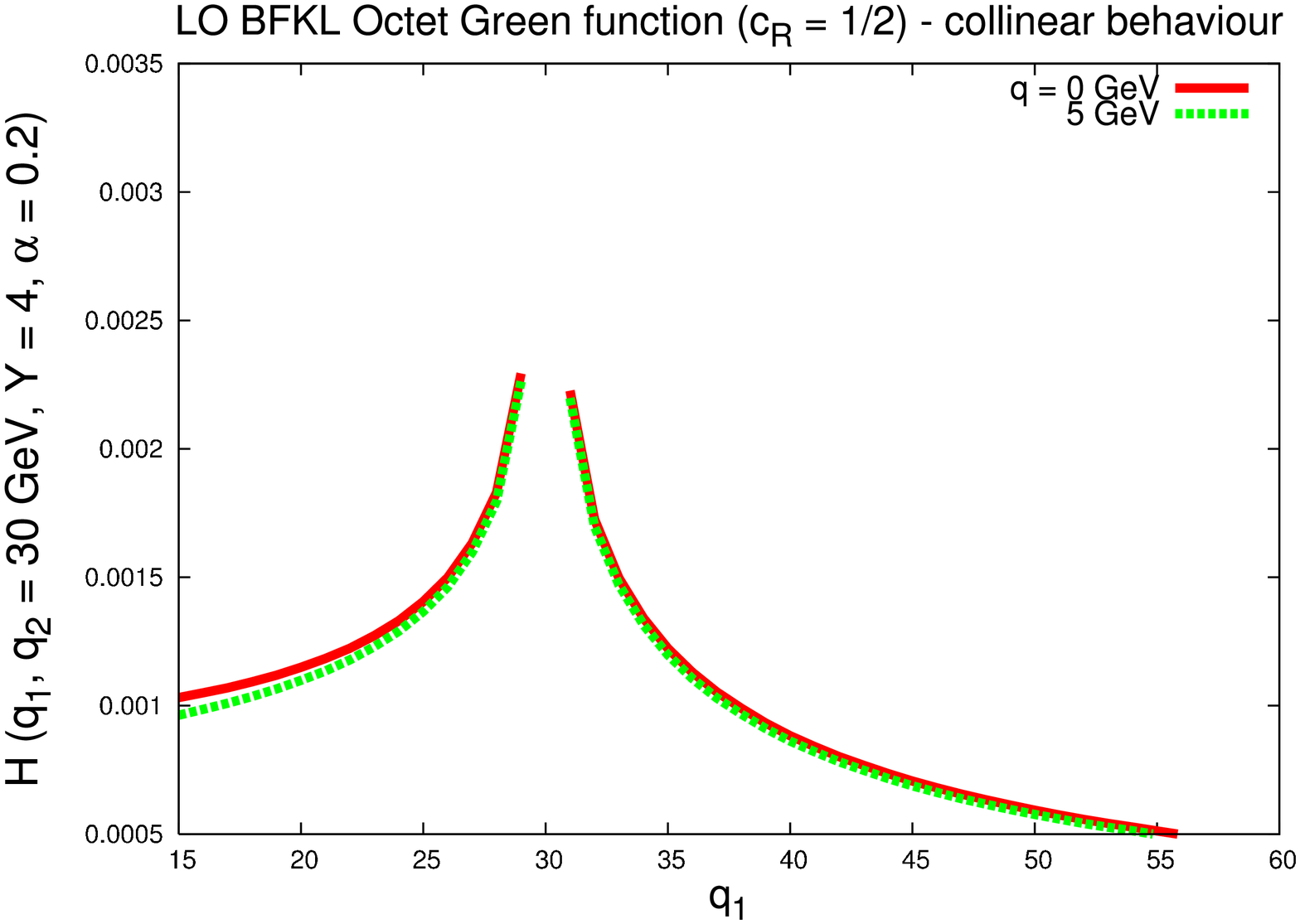}
\caption{Collinear behaviour of the gluon Green function.}
  \label{BFKLfcollinearso}
\end{figure}

\begin{figure}[htbp]
\hspace{3cm}\includegraphics[width=5cm,angle=-90]{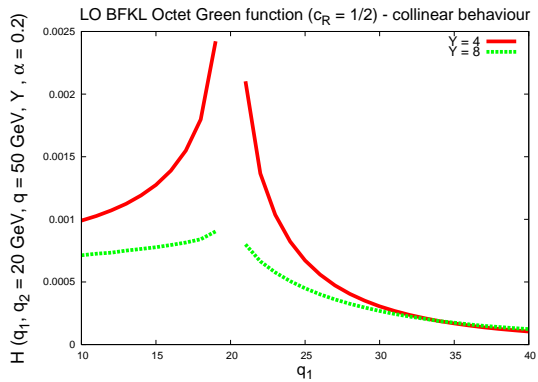}
\caption{Collinear behaviour of the octet gluon Green function at large momentum transfer.}
  \label{UsOctetColl}
\end{figure}

The most striking difference between the singlet and octet cases appears in the collinear limit. This is 
investigated in Fig.~\ref{BFKLfcollinearso}. In the region with low transverse momenta the octet Green function 
gets a big decrease when compared to the singlet one. This indicates that the corresponding anomalous dimensions governing this limit are quite different in both cases.
The collinear behaviour for the octet representation at large momentum transverse is shown in Fig.~\ref{UsOctetColl}

\section{Conformally invariant representations}

As it was shown in~\cite{Bartels:2008ce,Bartels:2008sc}, in order to have an explicitly $SL(2,C)$ invariant representation of the gluon Green function at large momentum transfer it is possible to subtract in Eq.~(\ref{BFKL1}) the gluon Regge 
trajectory depending on the total momentum transfer ${\bf q}^2$. To translate that representation to our 
present notation we need to work with the following modified Green function ${{\cal J}}$:
\begin{eqnarray}
{\cal J} \left({\bf q}_1,{\bf q}_2;{\bf q};{\rm Y}\right) &=& 
\left(\left({\lambda^{2 } \over {\bf q}^2} \right)^{2 c_{\cal R}} \frac{\left({\bf q}^2\right)^2}{{\bf q}_1^2 {\bf q}_1'^2} \right)^{\frac{{\bar \alpha}_s}{2} {\rm Y} } 
\Bigg\{\delta^{(2)} \left({\bf q}_1-{\bf q}_2\right) \nonumber\\&&\hspace{-3.2cm}+\sum_{n=1}^\infty \prod_{i=1}^n c_{\cal R} \int 
{d^2 {\bf k}_i \over \pi {\bf k}_i^2} \theta ({\bf k}_i^2-\lambda^2) 
\,  {{\bar \alpha}_s \over 2} 
\Bigg(1+{\left({\bf q}_1'+\sum_{l=1}^{i-1}{\bf k}_l\right)^2 ({\bf q}_1+\sum_{l=1}^{i}{\bf k}_l)^2 - {\bf q}^2 {\bf k}_i^2 \over 
({\bf q}_1'+\sum_{l=1}^{i}{\bf k}_l)^2 \left({\bf q}_1+\sum_{l=1}^{i-1}{\bf k}_l\right)^2}\Bigg)\nonumber\\
&&\hspace{-3cm}\times
\int_0^{y_{i-1}} \hspace{-0.4cm} d y_i \left(\left({\bf q}_1+\sum_{l=1}^{i-1}{\bf k}_l\right)^2
\left({\bf q}_1'+\sum_{l=1}^{i-1}{\bf k}_l\right)^2 \over 
\left({\bf q}_1+\sum_{l=1}^{i}{\bf k}_l\right)^2
\left({\bf q}_1'+\sum_{l=1}^{i}{\bf k}_l\right)^2 \right)^{1+{{\bar \alpha}_s \over 2}y_i}\hspace{-0.6cm} \delta^{(2)} \left({\bf q}_1+\sum_{l=1}^n {\bf k}_l-{\bf q}_2\right) \Bigg\}.
\label{L}
\end{eqnarray}
The treatment of the infrared divergencies is equivalent in Eq.~(\ref{H}) and Eq.~(\ref{L}) since they carry the same multiplicative 
factor $\lambda^{2 c_{\cal R} {\bar \alpha}_s Y}$. They are different in a relative normalization of the form 
$\left(\sqrt{{\bf q}_1^2 {\bf q}_1'^2} / {\bf q}^2 \right)^{(1-c_{\cal R}) {\bar \alpha}_s  {\rm Y} } $ which is not relevant in the singlet case but 
diverges when ${\bf q}^2 \to 0$ for the octet representation.  In the ``real emission" 
part the modification with respect to Eq.~(\ref{BFKL1}) is 
\begin{eqnarray} 
1+{{\bf q}_1'^2 ({\bf q}_1+{\bf k})^2 - {\bf q}^2 {\bf k}^2 \over ({\bf q}_1'+{\bf k})^2 {\bf q}_1^2} &\to&
\Bigg(1+{{\bf q}_1'^2 ({\bf q}_1+{\bf k})^2 - {\bf q}^2 {\bf k}^2 \over ({\bf q}_1'+{\bf k})^2 {\bf q}_1^2}\Bigg)
\frac{{\bf q}_1^2 {\bf q}_1'^2 }{({\bf q}_1+{\bf k})^2 ({\bf q}_1'+{\bf k})^2}.
\end{eqnarray}
\begin{figure}[htbp]
\hspace{-1.2cm} \includegraphics[width=8cm,angle=0]{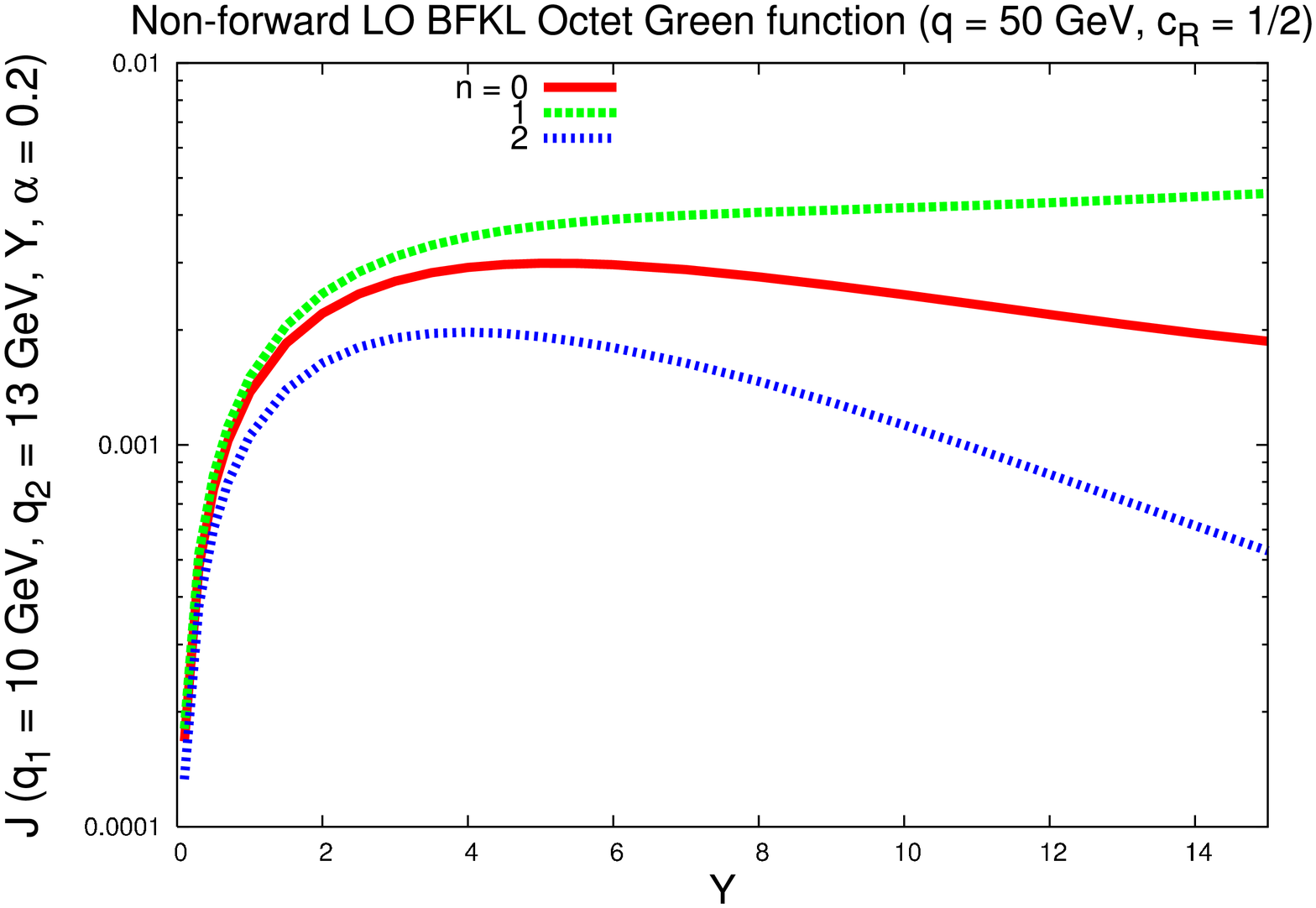}\includegraphics[width=8cm,angle=0]{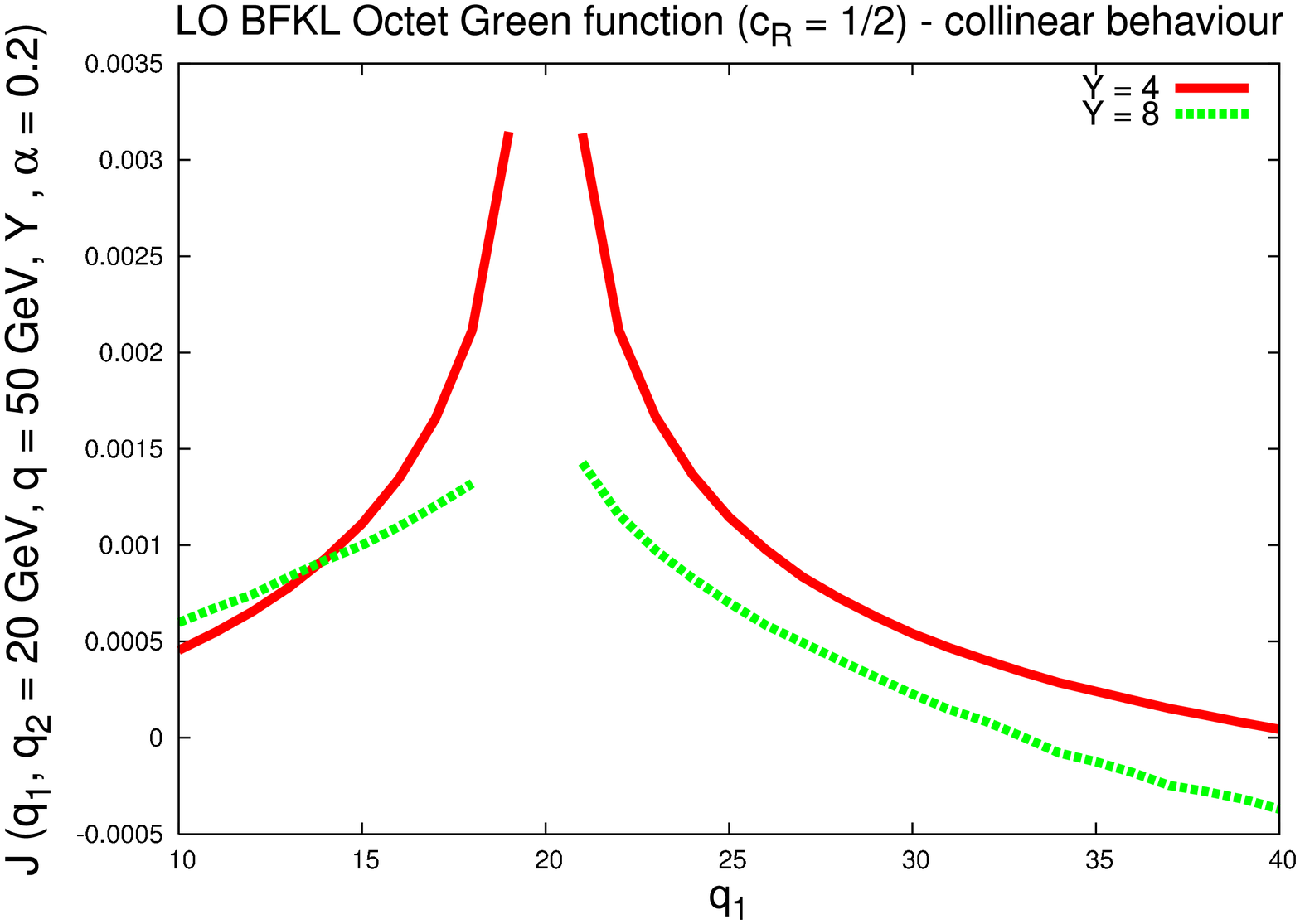}
\caption{Left: Projection of the function ${\cal J}$ on different conformal spins. Right: 
Collinear behaviour of the function ${\cal J}$ for two values of the rapidity $Y$.}
  \label{BFKLRepLev}  
\end{figure}
In the numerical analysis of the solution to this representation of the BFKL equation we find agreement with the work 
in~\cite{Bartels:2008ce,Bartels:2008sc} where it was shown that at large 
${\bf q}^2$ the dominant conformal spin is $n=1$, since it is the only one having a positive asymptotic intercept. 
The intercepts for all the other conformal spins, including $n=0$, are negative. We show this behaviour in Fig.~\ref{BFKLRepLev} (left) where we find that, in the limit ${\bf q}^2 \gg {\bf q}_1^2 \simeq {\bf q}_2^2$, the $n=1$ projection in the azimuthal angle is the only one growing with energy.

The collinear structure of the solution in the form of the function ${\cal J}$ is illustrated in Fig.~\ref{BFKLRepLev} (right) for two different values of $Y$. We find agreement with the one obtained in the previous section. It is 
also noteworthy the overall decrease of the Green function as $Y$ grows which occurs because in this plot we show the 
azimuthal angle integrated solution for which only the $n=0$ component survives.

\begin{figure}[htbp]
\hspace{-1.2cm} \includegraphics[width=8cm,angle=0]{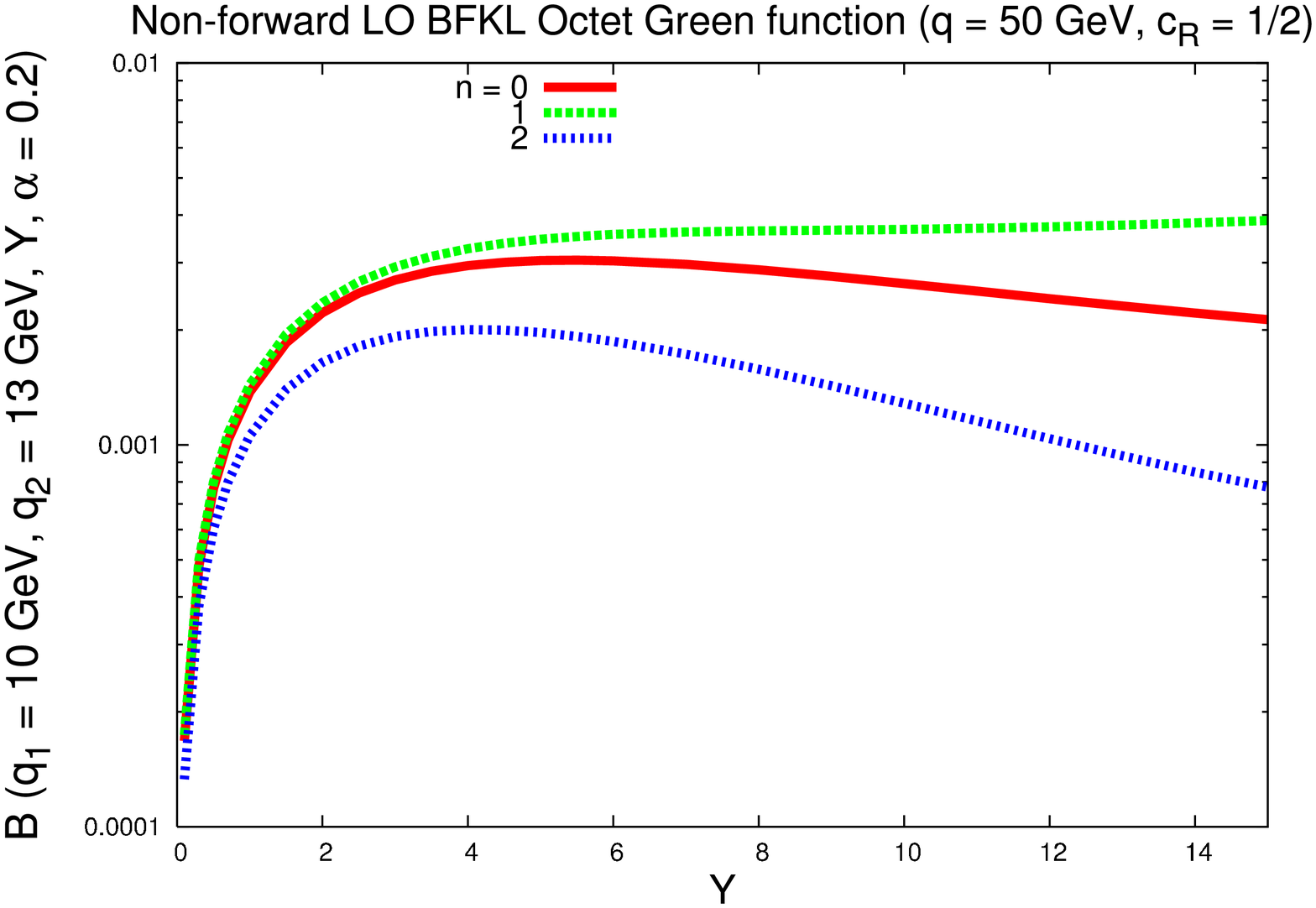}\includegraphics[width=8cm,angle=0]{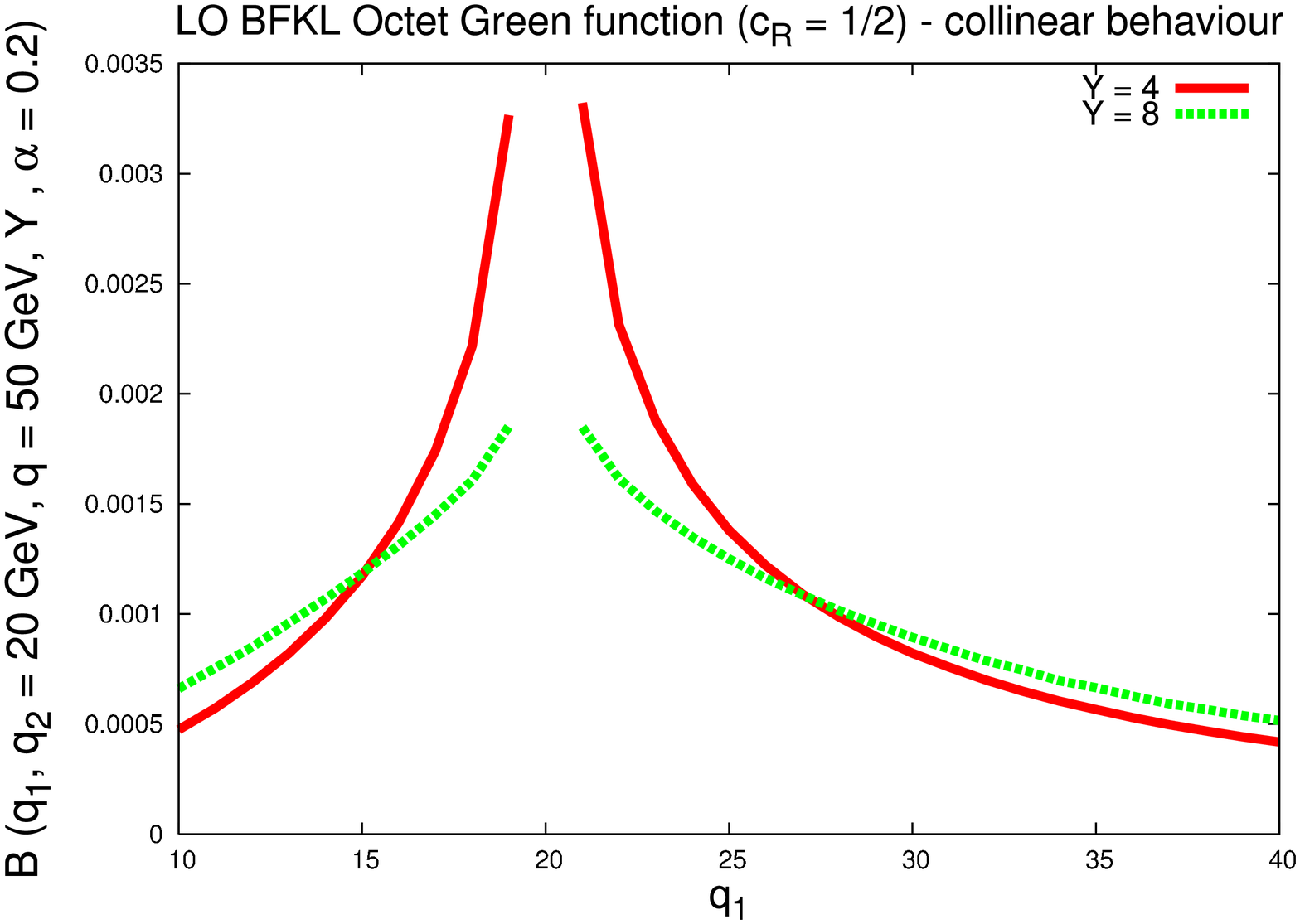}
\caption{Left: Projection of the function ${\cal B}$ on different conformal spins. Right: 
Collinear behaviour of the function ${\cal B}$ for two values of the rapidity $Y$.}
  \label{BFKLRepFL}  
\end{figure}

In a recent work Fadin and Lipatov~\cite{arXiv:1111.0782} have presented the NLO non-forward 
BFKL equation in the adjoint representation for the ${\cal N}=4$ supersymmetric theory. 
Their normalization for the Green function and treatment of the 
infrared divergencies, showing for simplicity only the LO pieces, is the following:
\begin{eqnarray}
\Bigg\{\omega + \frac{\bar \alpha}{2} 
\log{\left( {  {\bf q}_1^2 {\bf q}_1'^2   \over {\bf q}^2 \lambda^4}\right)}
\Bigg\} 
{\cal G}_\omega \left({\bf q}_1,{\bf q}_2;{\bf q}\right) &=&
\delta^{(2)} \left({\bf q}_1-{\bf q}_2\right)\nonumber\\
&&\hspace{-7cm}+ {{\bar \alpha}_s \over 4} \int {d^2 {\bf k} \over \pi {\bf k}^2} 
\theta \left({\bf k}^2 - \lambda^2\right)
\Bigg[1+{{\bf q}_1'^2 ({\bf q}_1+{\bf k})^2 - {\bf q}^2 {\bf k}^2 \over ({\bf q}_1'+{\bf k})^2 {\bf q}_1^2}\Bigg]
{{\bf q}_1^2 \over \left({\bf q}_1 + {\bf k}\right)^2}{\cal G}_\omega \left({\bf q}_1+{\bf k},{\bf q}_2;{\bf q}\right).
\end{eqnarray}
The corresponding solution in rapidity space reads:
\begin{eqnarray}
{\cal B} \left({\bf q}_1,{\bf q}_2;{\bf q};{\rm Y}\right) &=& 
\left({{\bf q}^2 \lambda^{2} \over {\bf q}_1^2 {\bf q}_1'^2 }
\right)^{\frac{\bar \alpha}{2}{\rm Y}} 
\Bigg\{\delta^{(2)} \left({\bf q}_1-{\bf q}_2\right) \nonumber\\
&&\hspace{-3.5cm}+\sum_{n=1}^\infty \prod_{i=1}^n \,  {{\bar \alpha} \over 4}  
\int {d^2 {\bf k}_i \over \pi {\bf k}_i^2} \theta ({\bf k}_i^2-\lambda^2) 
\Bigg(1+{\left({\bf q}_1'+\sum_{l=1}^{i-1}{\bf k}_l\right)^2 ({\bf q}_1+\sum_{l=1}^{i}{\bf k}_l)^2 - {\bf q}^2 {\bf k}_i^2 \over ({\bf q}_1'+\sum_{l=1}^{i}{\bf k}_l)^2 \left({\bf q}_1+\sum_{l=1}^{i-1}{\bf k}_l\right)^2}\Bigg)
\int_0^{y_{i-1}} d y_i  \nonumber\\
&&\hspace{-3cm} \times
\left(\left({\bf q}_1+\sum_{l=1}^{i-1}{\bf k}_l\right)^2 \over 
\left({\bf q}_1+\sum_{l=1}^{i}{\bf k}_l\right)^2 \right)^{1+ {{\bar \alpha} y_i \over 2}}
\left(\left({\bf q}_1'+\sum_{l=1}^{i-1}{\bf k}_l\right)^2 \over 
\left({\bf q}_1'+\sum_{l=1}^{i}{\bf k}_l\right)^2 \right)^{ {\bar \alpha} y_i \over 2}
\delta^{(2)} \left({\bf q}_1+\sum_{l=1}^n {\bf k}_l-{\bf q}_2\right) \Bigg\},
\end{eqnarray}
The difference with the ${\cal J}$ function is the normalization of the Green function, while the treatment of the infrared divergences 
is the same. It is then natural to obtain a very similar behaviour for ${\cal B}$ as it is shown in Fig.~\ref{BFKLRepFL}.

\section{Conclusions and scope}

In this short letter, a detailed study of the solution to the octet BFKL equation has been carried out. So far, 
this solution has only been discussed in~\cite{Bartels:2008ce,Bartels:2008sc,arXiv:1111.0782}  
where it naturally appears in the context of maximally 
helicity violating planar amplitudes in ${\cal N} = 4$ super Yang-Mills theory. In certain kinematical  regions, the gluon 
Green function in the octet sector plays a fundamental role to obtain the so-called ``finite remainder function". 
It is therefore very important to understand its structure in depth, not only to the first orders in the `t Hooft coupling, but also its all-orders features. This has been the target of the work presented here. We have discussed several representations directly written in the transverse momentum space which allow for the factorization of its infrared divergencies. The dominant Fourier component in the azimuthal angle is different in these representations, also 
showing a different collinear limit when compared to the singlet case. 
The integration of these representations with the corresponding impact factors to investigate complete 
scattering amplitudes to all orders in the coupling will be the subject of our future investigations. The study of the 
next-to-leading order color octet kernel is also underway. 
\\
\\
\\
{\bf \large Acknowledgements}\\
We would like to thank Jochen Bartels and Lev Lipatov for useful discussions. G. C. thanks the Department of Theoretical Physics at the Aut{\'o}noma University of Madrid and the ``Instituto de F{\'\i}sica Te{\' o}rica 
UAM / CSIC" for their hospitality. We thank the CERN PH-TH unit where part of this work was performed. 
Research partially supported by the European Comission under contract LHCPhenoNet (PITN-GA-2010-264564), 
the Comunidad de Madrid through Proyecto HEPHACOS ESP-1473, and MICINN (FPA2010-17747).

\end{document}